\documentstyle[12pt]{article}
\begin{document}
\baselineskip 10mm

\vskip 4mm

\centerline{\large \bf Localized electron state in a T-shaped confinement
potential}

\vskip 12mm

\centerline{L. A. Openov}

\vskip 2mm

\centerline{\it Moscow State Engineering Physics Institute
(Technical University),}
\centerline{\it Kashirskoe sh. 31, Moscow 115409, Russia}

\vskip 20mm

\centerline{\bf Abstract}

\begin{quotation}

We consider a simple model of an electron moving in a T-shaped confinement
potential. This model allows for an analytical solution that explicitly
demonstrates the existence of laterally bound electron states in quantum
wires obtained by the cleaved edge overgrowth technique.

\end{quotation}

\newpage

In recent years, semiconductor nanostructures have attracted much attention
due to both a great variety of their possible applications in electronic
devices and a new interesting physics emerged to describe their peculiar
characteristics \cite{meso}. Among other research directions, there are
intensive studies of the so called quantum wires \cite{wires}, in which
electrons are confined in two spatial dimensions within a nanometer-size
region, while being allowed to move freely in the third direction (an axis
of the wire).

High-quality quantum wires can be fabricated by the molecular-beam epitaxy
using the cleaved edge overgrowth (CEO) technique \cite{Pfeiffer}, as
proposed by Esaki {\it et al.} \cite{Esaki}. Such structures are formed at
the T-shaped intersection of two semiconductor quantum wells, see fig. 1. The
electronic band structure of realistic T-shaped AlGaAs/GaAs quantum wires has
been studied numerically using different methods \cite{theory}.

Note, however, that the form of the T-shaped confinement potential, fig. 1,
does not allow for analytical solutions of the one-electron continuum
Schr\"odinger equation even in the limiting case of infinite energy barriers
and isotropic electron effective mass. Meanwhile, the existence of electron
states localized in the region of intersection of two quantum wells within
the plane perpendicular to the axis of CEO quantum wire is not obvious
{\it a priori}. In this paper, we propose a simple model of the
two-dimensional T-shaped structure and obtain an analytical solution for both
the lowest energy eigenvalue and corresponding localized eigenfunction of the
one-electron Schr\"odinger equation.

Let us consider two intersecting atomic chains, one of which (along
{\it x}-axis) is infinite, while another (along {\it y}-axis) is
semi-infinite, see fig. 2. Such a system provides a model for the description
of electron motion in the {\it xy}-plane perpendicular to the axis of a
CEO quantum wire (fig. 1) since an electron is confined to the chains and
can not escape into the interjacent region. This model corresponds to the
limiting case of a CEO quantum wire formed by two quantum wells
with one-atomic-layer width and an infinite value of the energy barrier
$V=E_c^A-E_c^B$ preventing an electron from going to the conduction band of
the semiconductor A, see fig. 1.

In a tight-binding approach, one-electron states of the system under
consideration are described by the following Hamiltonian:
\begin{equation}
\hat{H}(t)=
\epsilon_0\sum_{i}{\hat{a}^{+}_{i}\hat{a}^{}_{i}}-
t\sum_{<i,j>}{(\hat{a}^{+}_{i}\hat{a}^{}_{j}+h.c)}~,
\label{Ham}
\end{equation}
where $\epsilon_0$ is an on-site potential, $t$ is a matrix element for
electron hopping between nearest sites $i$ and $j$, $\hat{a}^{+}_{i}$
$(\hat{a}^{}_{i})$ is the operator of electron creation (annihilation) at the
site $i$, and $<i,j>$ means the summation over nearest neighbors only. We do
not specify the spin index explicitly since we consider the states of a
single electron.

Expanding the one-electron wave function $\Psi$ into atomic electron states
$|i>$,
\begin{equation}
\Psi=\sum_{i}A_i|i>~,
\label{Psi}
\end{equation}
we have from the Schr\"odinger equation $\hat{H}\Psi=E\Psi$ a set of
algebraic equations for coefficients $A_i$ whose squared absolute values
$|A_i|^2$ give the probabilities to find an electron at a particular site
$i$:
\begin{equation}
\tilde{E}A_i=\sum_{j=nn(i)} A_j~,
\label{Ai}
\end{equation}
where
\begin{equation}
\tilde{E}=\frac{\epsilon_0-E}{t}
\label{Etilde}
\end{equation}
and $j=nn(i)$ means the summation over sites $j$ nearest to the site $i$.

Let $i=n$ for atoms in the chain running along {\it x}-axis and $i=m$ for
atoms in the chain along {\it y}-axis ($n = 0, \pm 1, \pm 2, \pm 3,... ;
m = 0, 1, 2, 3, ...$), see fig. 2. For the atom belonging to both chains (at
the intersection point) one has $n=m=0$. We consider three different regions:
(1) $n \ge 0$; (2) $n \le 0$; (3) $m \ge 0$. In those regions, we have from
eq. (\ref{Ai}):
\begin{eqnarray}
\tilde{E}A_n=A_{n-1}+A_{n+1}~,~n\ge 1~,
\nonumber \\
\tilde{E}A_n=A_{n-1}+A_{n+1}~,~n\le -1~,
\nonumber \\
\tilde{E}A_m=A_{m-1}+A_{m+1}~, m\ge 1~.
\label{Anm}
\end{eqnarray}
General solutions of eqs. (\ref{Anm}) are:
\begin{eqnarray}
A_n = A_0 \exp(-\alpha n)~,~n\ge 0~,
\nonumber \\
A_n = B_0 \exp(\beta n)~,~n\le 0~,
\nonumber \\
A_m = C_0 \exp(-\gamma n)~,~m\ge 0~,
\label{Anm1}
\end{eqnarray}
where
\begin{equation}
\tilde{E}=2\cosh(\alpha)=2\cosh(\beta)=2\cosh(\gamma)~.
\label{Etilde1}
\end{equation}
Since, due to normalization condition $\sum_{i}|A_i|^2=1$, the coefficients
$A_n$ and $A_m$ should be finite at $n\rightarrow \pm \infty$ and
$m\rightarrow \infty$ respectively, one has
$Re(\alpha) \ge 0,~Re(\beta) \ge 0,~Re(\gamma) \ge 0$.
Hence, we see from eq. (\ref{Etilde1}) that $\alpha=\beta=\gamma$, while from
eqs. (\ref{Anm1}) we have $A_0=B_0=C_0$ since $A_{n=0}=A_{m=0}$ at the
intersection point.

Next, from eq. (\ref{Ai}) for the coefficient $A_0$ we obtain
\begin{equation}
\tilde{E}A_0=A_{n=1}+A_{n=-1}+A_{m=1}~.
\label{A0}
\end{equation}
Taking eqs. (\ref{Anm1}) into account, we have
\begin{equation}
\tilde{E}=3\exp(-\alpha)~.
\label{Etilde2}
\end{equation}
The lowest energy solution of eqs. (\ref{Etilde1}) and (\ref{Etilde2}) is
$\tilde{E}=3/\sqrt{2}$ and $\exp(-\alpha)=1/\sqrt{2}$, and so from
eq. (\ref{Etilde}) we have the minimum one-electron energy of the system
under consideration:
\begin{equation}
E_{min}=\epsilon_0-\frac{3}{\sqrt{2}}t~,
\label{E}
\end{equation}
while the corresponding normalized wave function is
\begin{equation}
\Psi=\sum_{n=-\infty}^{\infty}\frac{1}{2^{|n|/2+1}}|n>+
\sum_{m=1}^{\infty}\frac{1}{2^{m/2+1}}|m>~.
\label{Psi1}
\end{equation}
One can see from eq. (\ref{Psi1}) that the wave function is localized
exponentially in the vicinity of the intersection point $n=0, m=0$. The
probability to find an electron at the site $n=m=0$ equals to 1/4, while
the probabilities to find an electron at the nearest sites $n=-1, n=1, m=1$
are 1/8 each. The localization length $\xi$ defined by $A_n=\exp(-a|n|/\xi)$,
where $a$ is the interatomic spacing, equals to
$\xi=a/\alpha=2a/\ln(2)\approx 3a$.

We note that the value of $E_{min}=\epsilon_0-3t/\sqrt{2}$ is by
$\approx 0.12t$ lower than the edge $E{_0}=\epsilon_0-2t$ of
the band of delocalized electron states in the infinite one-dimensional
chain. An estimate of the hopping matrix element $t$ for a specific case of
GaAs gives in a tight-binding approach $t=\hbar^2/2m^*a^2\approx$
1.8 eV, where $m^*=0.067m_0$, $a=$ 0.565 nm, $m_0$ is the mass of a free
electron. Then for the confinement energy one has
$E_c=E{_0}-E_{min}\approx$ 200 meV. The physical reason for appearance of
the localized state in the T-shaped geometry seems to be related with the
effect of Anderson localization \cite{Anderson} since the semi-infinite
chain along $y$-axis plays a role of defect for an electron moving in the
infinite chain along $x$-axis.

We have also studied the model under consideration by the exact numerical
diagonalization of the Hamiltonian matrix. We made use of the complex
conjugate method which allows one to find eigenvalues and eigenvectors of
large sparse matrixes with any prescribed accuracy \cite{Openov}. For our
purposes, we have restricted ourselves to several low-lying levels. We have
considered the systems composed of $3N+1$ sites ($2N+1$ sites in the chain
along $x$-axis and $N+1$ sites in the chain along $y$-axis, one site being
common for both chains), with $N$ up to $10^3$. For the ground state, the
calculated values of $E_{min}$ and $\alpha$ coincide with analytical results.
The excited states have the energies $E_i\ge E{_0}$, where the value of
$E{_0}$ tends to $\epsilon_0-2t$ as $N$ increases, e.g.,
$E{_0}=\epsilon_0-1.99903t$ for $N=$ 100 and $E{_0}=\epsilon_0-1.99976t$
for $N=$ 200. The mean level spacing of the excited states is of the order of
$t/N$ and goes to zero as $N$ increases. An inspection of wave functions of
low-lying excited states has shown that all of them are delocalized over the
whole system, having the form of sines or cosines.

Finally, along with the system shown in fig. 2, we have examined a more
general case of quasi-one-dimensional chains of finite width, so that each
chain had $N_0\ge 2$ sites in width ($N_0=$ 1 in a particular case studied
analytically above). We have considered the systems composed of
$3N\cdot N_0+N_0^2$ sites ($(2N+N_0)\cdot N_0$ sites in the chain along
$x$-axis and $(N+N_0)\cdot N_0$ sites in the chain along $y$-axis, $N_0^2$
sites being common for both chains), with $N$ up to $10^4$ and $N_0$ up to
40. We have found that the ground state energy $E_{min}$ decreases with
increasing $N_0$, see fig. 3, and remains lower than the edge of the band of
delocalized states, $E{_0}$. Fig. 4 shows the confinement
energy $E_c=E{_0}-E_{min}$ as a function of the chain width $N_0$. One can
see that $E_c$ decreases with $N_0$ and tends to zero as $N_0$ goes to
infinity. Taking the values of $t$ and $a$ for GaAs, see above, one has,
e.g., $E_c\approx 32$ meV and 9 meV for the chains of 5 nm and 10 nm width
respectively. These values of $E_c$ are about twice as large as those
obtained for the conduction band of symmetrical T-shape wire structures with
GaAs wells and Al$_{0.3}$Ga$_{0.7}$As barriers within a much more
sophisticated model \cite{theory}. An apparent reason for this quantitative
discrepancy is the infinite value of the conduction-band offset in our simple
model. So, our approach provides an upper estimate for the electron
confinement energy in the T-shaped quantum wires.

From figs. 3 and 4 we find that at large $N_0$ both $E_{min}$ and
$E{_0}$ approach -4$t$ which is just the value of $E{_0}=E_{min}$ for the
infinite two-dimensional system in a tight-binding model. Again, the ground
state wave function is always
localized exponentially in the vicinity of the intersection region, while the
wave functions of excited states are delocalized. The values of the
localization length $\xi$ at different $N_0$ where found through the analysis
of the asymptotic behavior of the ground state wave function at large
distance from the intersection region. It was found that the decrease in
$E_c$ with $N_0$ is accompanied by the corresponding increase in $\xi$
according to the general relation $E_c=\hbar^2/2m^*\xi^2=t(a/\xi)^2$.
The dependence of $\xi$ on $N_0$ is shown in fig. 5.

In conclusion, we have presented an analytically solvable model of the
T-shaped two-dimensional confinement potential. Although being rather simple,
this textbook model explicitly demonstrates the existence of localized
electron states in the T-shaped geometry. It can be used for qualitative
estimates of confinement energies and the extent of spatial localization
of one-electron wave function in CEO quantum wires. The model can be easily
generalized to account for electron hopping to atoms other than nearest
neighbors only and/or for the finite value of the confinement potential. In
such a case, however, it will be difficult to obtain an analytical solution
of one-electron Schr\"odinger equation.

\vskip 4mm

This work was supported in part by the Russian Federal Program "Integration",
project No A0133.

\newpage

\centerline{\bf FIGURE CAPTIONS}

\vskip 2mm

Fig. 1. Schematic view of a T-shaped semiconductor quantum wire formed by
two intersecting quantum wells. The height of the energy barrier for
electrons in the semiconductor B equals to the conduction-band offset
$\Delta E_c=E_c^A-E_c^B$ between semiconductors A and B. Electrons are
confined in the $xy$-plane in the vicinity of intersection of quantum wells
and move freely along $z$-axis (an axis of the quantum wire)

Fig. 2. Two intersecting atomic chains as a model of a T-shaped confinement
potential for electrons in the plane perpendicular to the axis of CEO quantum
wire. Numbers $n$ and $m$ numerate atoms in the chains along $x$- and
$y$-axis respectively.

Fig. 3. The energy $E_{min}$ of the localized state in units of the hopping
matrix element $t$ versus the number of sites $N_0$ across the chain. Points
are the results of analytical (for $N_0=$ 1) and numerical (for $N_0\ge 2$,
$N=$ 800) calculations. The line is a guide to the eye.

Fig. 4. The confinement energy $E_c=E{_0}-E_{min}$ in units of the hopping
matrix element $t$ versus the number of sites $N_0$ across the chain. Points
are the results of analytical (for $N_0=$ 1) and numerical (for $N_0\ge 2$,
$N=$ 800) calculations. The line is a guide to the eye.

Fig. 5. Localization length $\xi$ of the ground state wave function in
units of the interatomic spacing $a$ versus the number of sites $N_0$ across
the chain. Points are the results of analytical (for $N_0=$ 1) and numerical
(for $N_0\ge 2$, $N=$ 800) calculations. The line is the curve
$\xi=a(t/E_c)^{1/2}$, where $E_c=E{_0}-E_{min}$ is the confinement energy.

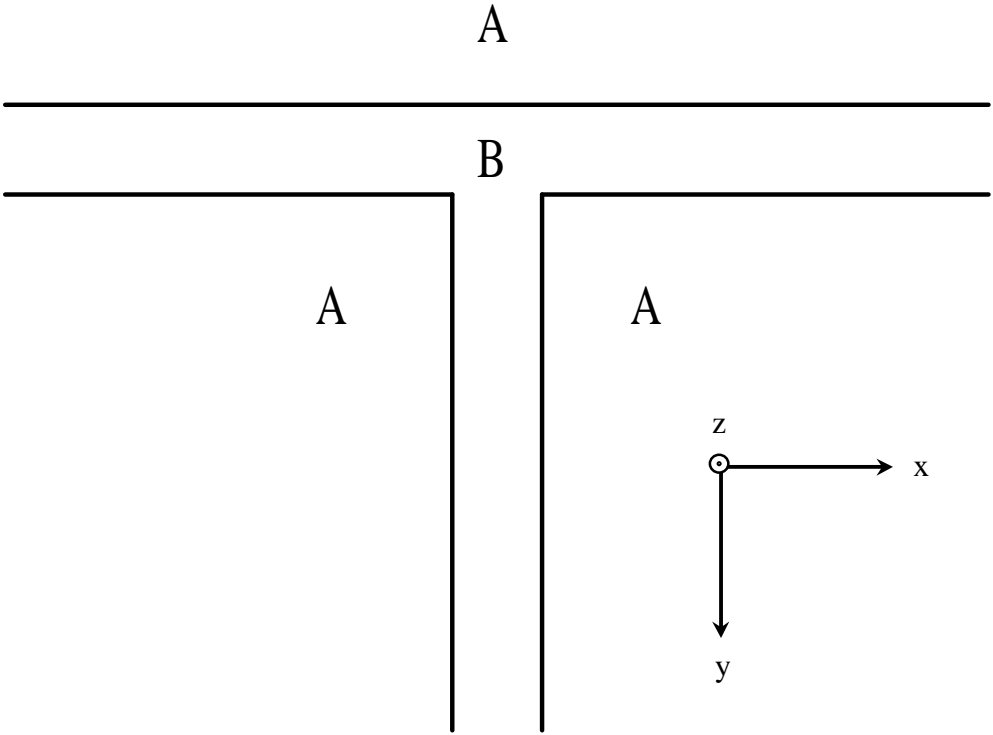
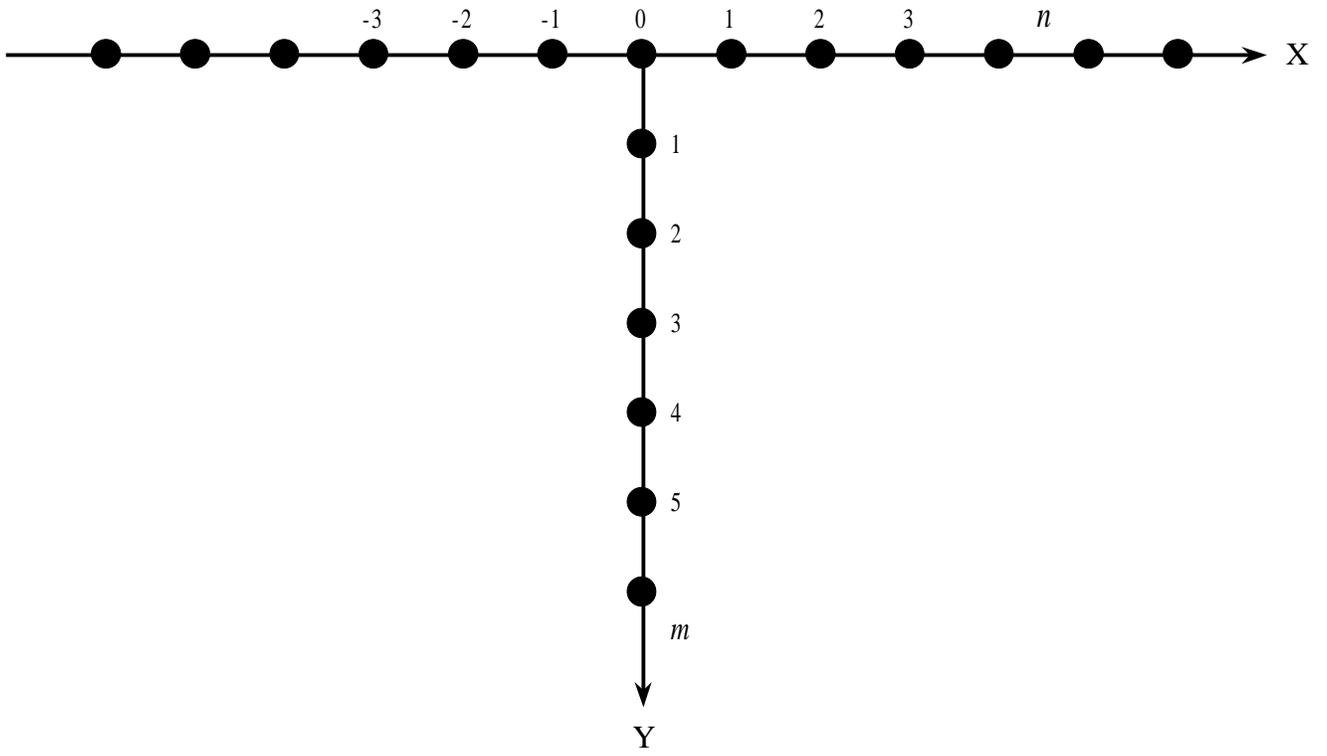
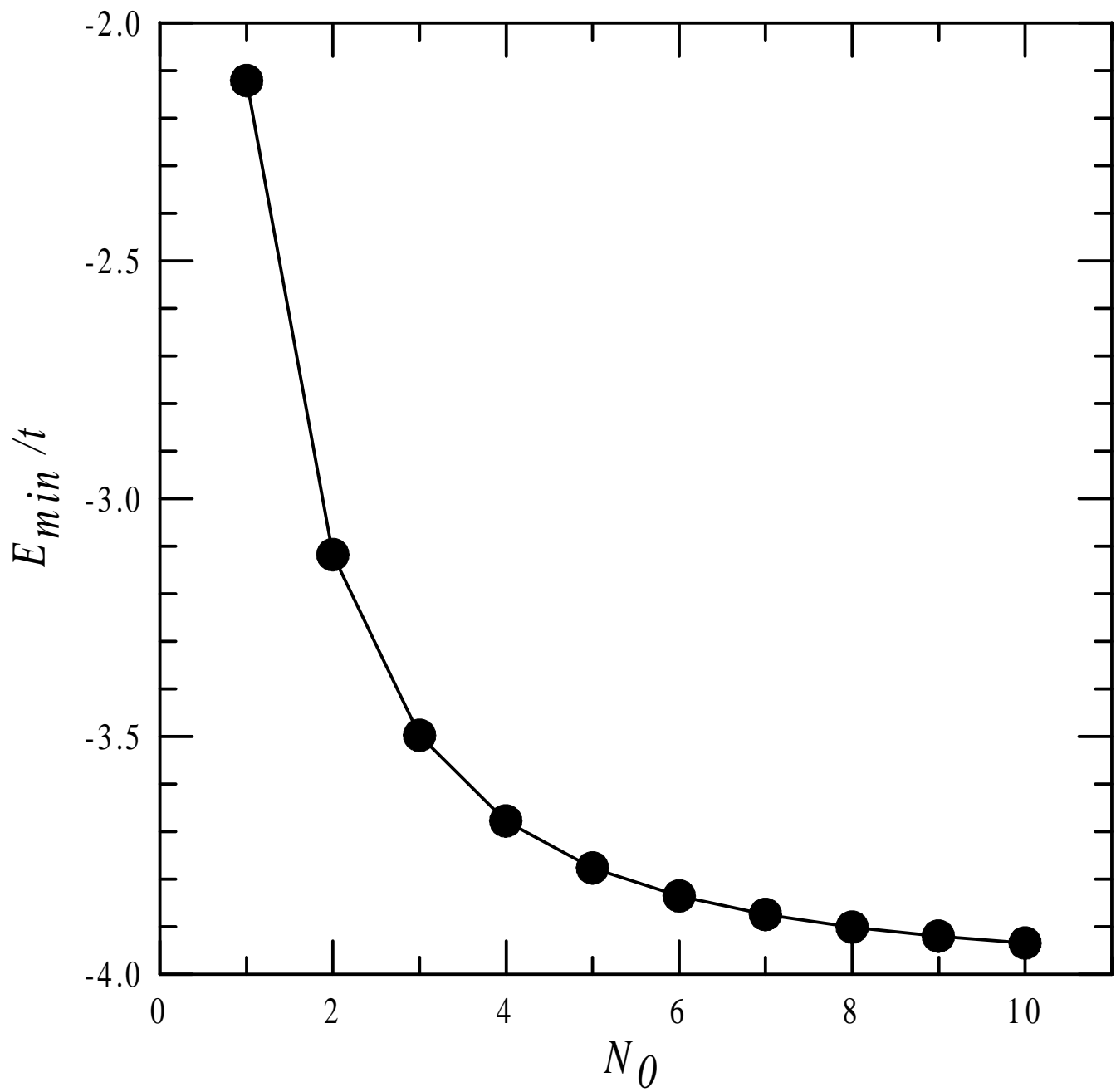
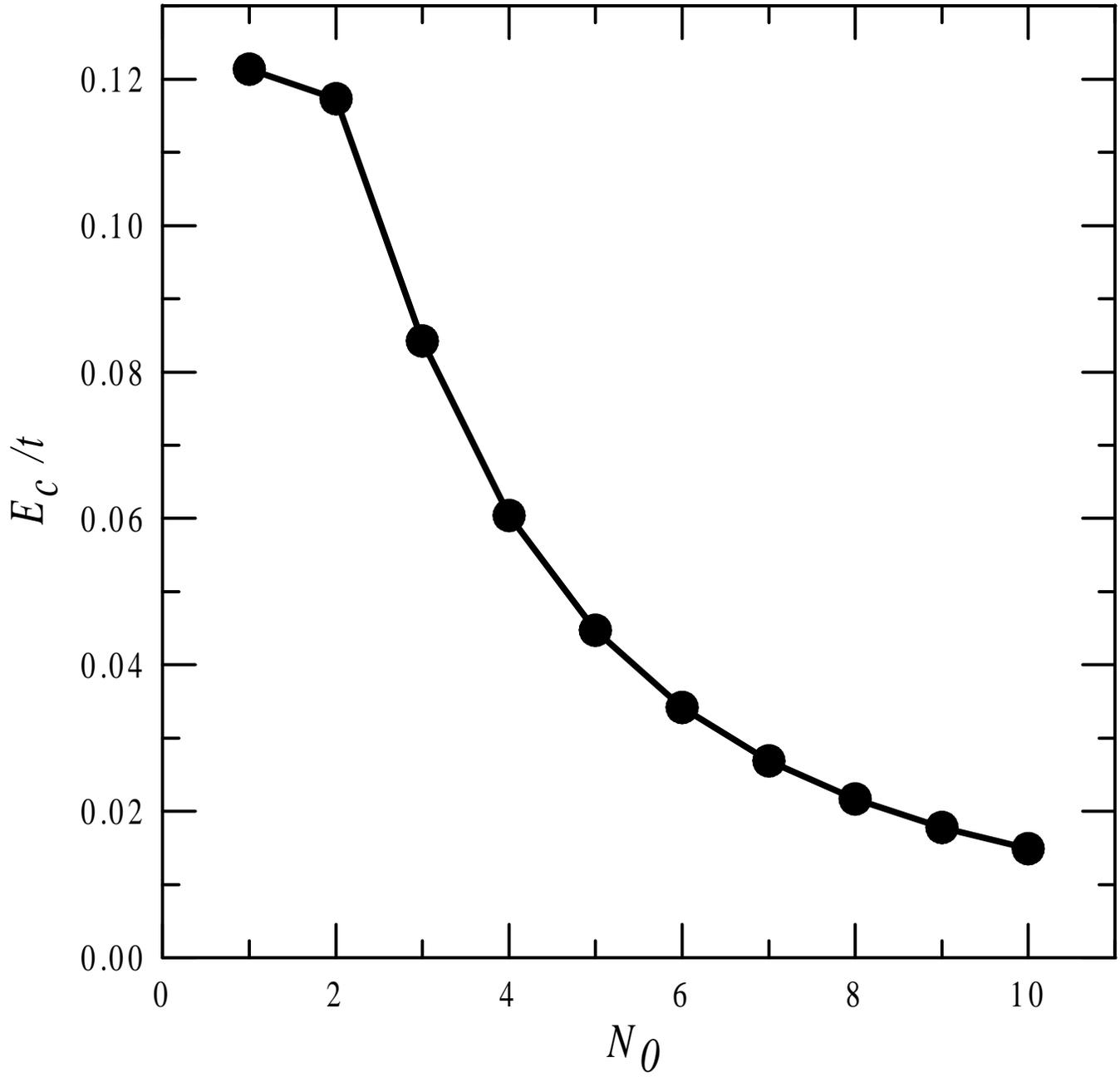
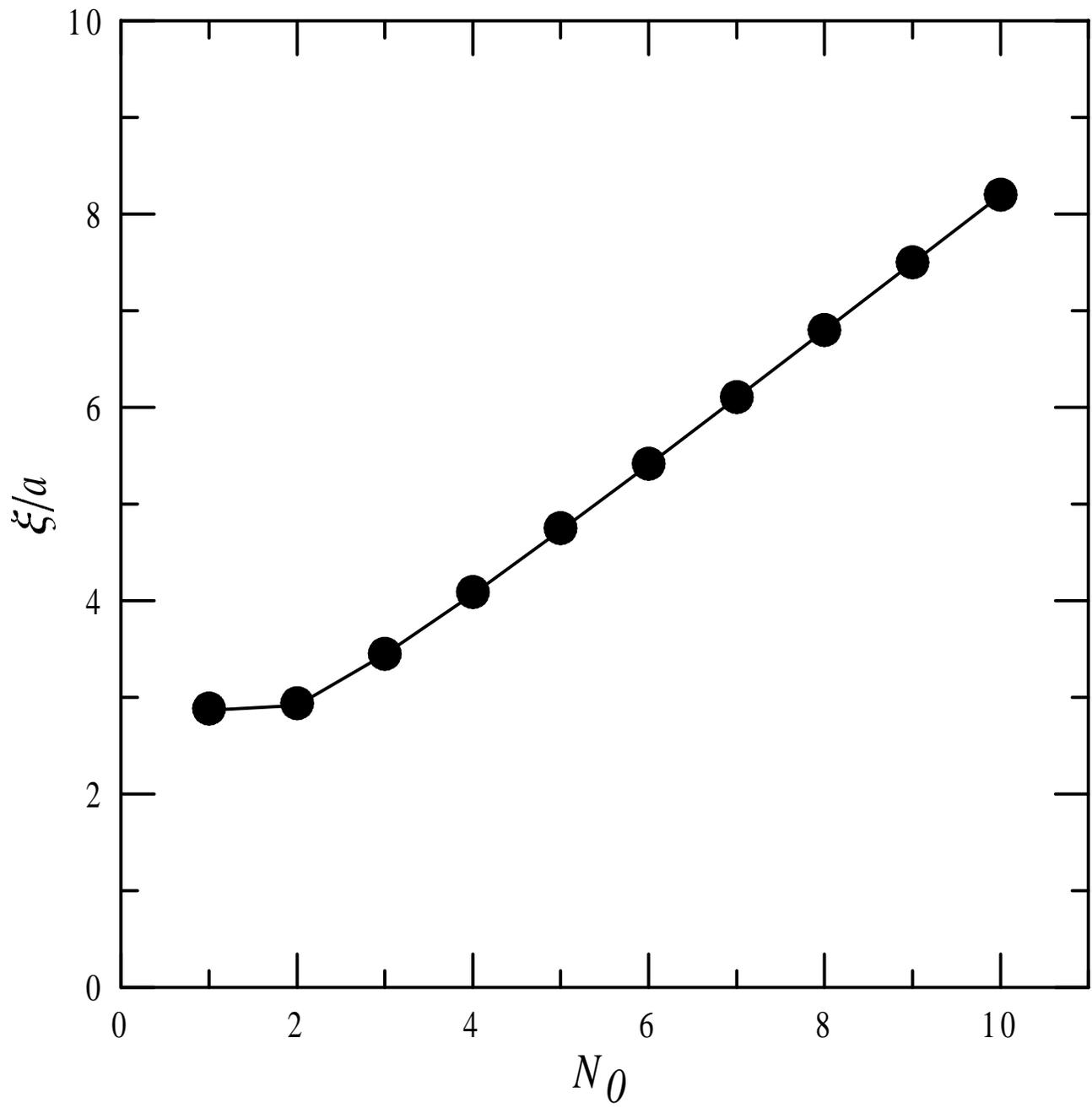
\end{document}